\documentclass[draft,jgrga]{AGUTeX}

\usepackage{amsmath}

\usepackage{graphicx}

\setkeys{Gin}{draft=false}

\authorrunninghead{OROSEI ET AL.}

\titlerunninghead{RADAR SOUNDING OF LUCUS PLANUM}

\authoraddr{Corresponding author: Roberto Orosei, Istituto di Radioastronomia, Istituto Nazionale di Astrofisica, Bologna, Italy. (roberto.orosei@inaf.it)}

\begin{document}


\title{Radar sounding of Lucus Planum, Mars, by MARSIS}




\authors{Roberto Orosei,\altaffilmark{1}
Angelo Pio Rossi,\altaffilmark{2}
Federico Cantini,\altaffilmark{3}
Graziella Caprarelli,\altaffilmark{4,5}
Lynn M. Carter,\altaffilmark{6}
Irene Papiano,\altaffilmark{7}
Marco Cartacci,\altaffilmark{8}
Andrea Cicchetti,\altaffilmark{8}, and
Raffaella Noschese\altaffilmark{8}}

\altaffiltext{1}{Istituto Nazionale di Astrofisica, Istituto di Radioastronomia, Via Piero Gobetti 101, 40129 Bologna, Italy}
\altaffiltext{2}{Department of Physics and Earth Sciences, Jacobs University Bremen, Campus Ring 1, 28759 Bremen, Germany}
\altaffiltext{3}{Ecole Polytechnique Federale de Lausanne, Space Engineering Center, EPFL ESC, Station 13, 1015 Lausanne, Switzerland}
\altaffiltext{4}{University of South Australia, Div ITEE, GPO Box 2471, Adelaide SA 5001, Australia}
\altaffiltext{5}{International Research School of Planetary Sciences, Viale Pindaro 42, Pescara 65127, Italy}
\altaffiltext{6}{The University of Arizona, Lunar and Planetary Laboratory, 1629 E University Blvd, Tucson, AZ 85721-0092, USA}
\altaffiltext{7}{Liceo Scientifico Augusto Righi, Viale Carlo Pepoli 3, 40123 Bologna, Italy}
\altaffiltext{8}{Istituto Nazionale di Astrofisica, Istituto di Astrofisica e Planetologia Spaziali, Via del Fosso del Cavaliere 100, 00133 Roma, Italy}



\begin{abstract}
Lucus Planum, extending for a radius of approximately 500 km around $181^{\circ}$ E, $5^{\circ}$ S, is part of the Medusae Fossae Formation (MFF), a set of several discontinuous deposits of fine-grained, friable material straddling across the Martian highland-lowland boundary. The MFF has been variously hypothesized to consist of pyroclastic flows, pyroclastic airfall, paleopolar deposits, or atmospherically-deposited icy dust driven by climate cycles. MARSIS, a low--frequency subsurface--sounding radar carried by ESA's Mars Express, acquired 238 radar swaths across Lucus Planum, providing sufficient coverage for the study of its internal structure and dielectric properties. Subsurface reflections were found only in three areas, marked by a distinctive surface morphology, while the central part of Lucus Planum appears to be made of radar--attenuating material preventing the detection of basal echoes. The bulk dielectric properties of these areas were estimated and compared with those of volcanic rocks and ice--dust mixtures. Previous interpretations that east Lucus Planum and the deposits on the north--western flanks of Apollinaris Patera consist of high--porosity pyroclastic material are strongly supported by the new results. The north--western part of Lucus Planum is likely to be much less porous, although interpretations about the nature of the subsurface materials are not conclusive. The exact origin of the deposits cannot be constrained by radar data alone, but our results for east Lucus Planum are consistent with an overall pyroclastic origin, likely linked to Tharsis Hesperian and Amazonian activity.
\end{abstract}



\begin{article}

\section{Introduction}

Lucus Planum, extending for a radius of approximately 500 km around $181^{\circ}$ E, $5^{\circ}$ S, is part of the Medusae Fossae Formation (MFF), a set of several discontinuous deposits of fine-grained, friable material straddling across the Martian highland-lowland boundary (e.g. \citep[e.g.][]{2009Icar..199..295C}).

\begin{figure}
\includegraphics[width=0.9\linewidth]{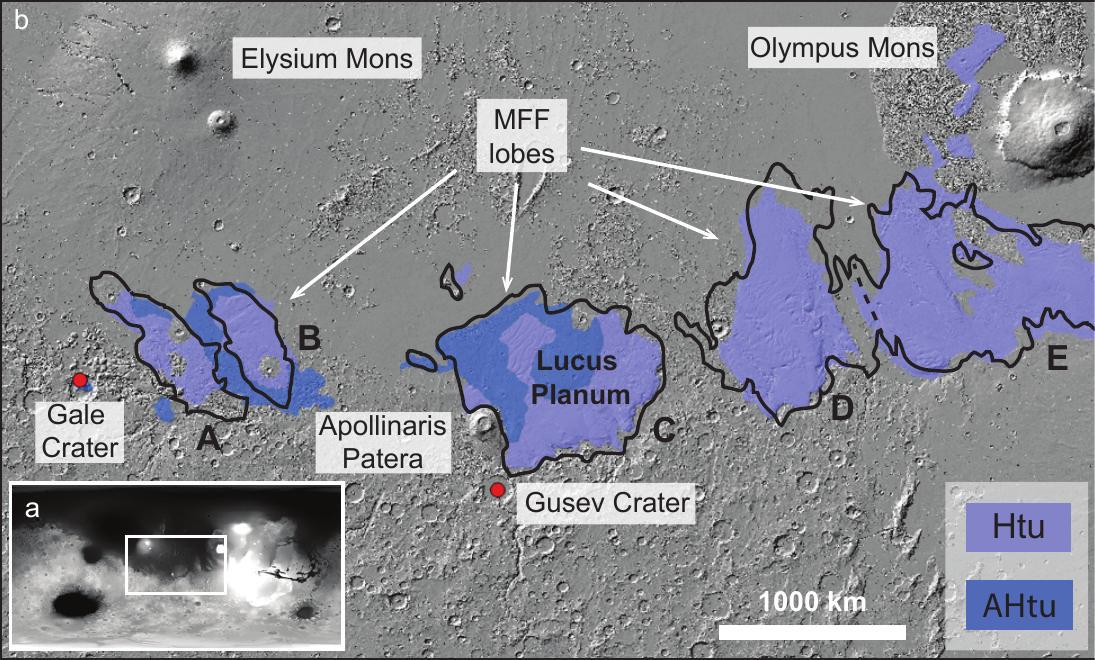}
\caption{Geomorphic and geologic setting for Lucus Planum: a) Location of the greater Medusa Fossae Formation (MFF) deposits, the sub-figure b is outlined in white. b) Extent of the Medusa Fossae Formation as outlined, in black, by \citet{2011Icar..216..212K}. The MFF lobes are also indicated with letters, according to \citet{harrison2010mapping}. Lucus Planum constitutes the centre lobe of the MFF. Units from the revised global geological map of Mars \citep{tanaka2014geologic} are indicated {\it Htu} and {\it AHtu}. The location of the NASA MER Spirit rover landing Site in Gusev crater and NASA MSL Curiosity in Gale crater are indicated. Figure 1a is a global map in equirectangular projection, centred at longitude 180. Figure 1b is a map centred on Lucus Planum and covering the MFF, in equirectangular projection, and centred at $180^{\circ}$ E. The longitudinal extension is about 5000 km. Latitude range is from $-30^{\circ}$ to $30^{\circ}$, and longitude ranges from $130^{\circ}$ E to $230^{\circ}$ E.}
\label{fig01new}
\end{figure}

The MFF covers an extensive area, spanning latitudinally more than 1000 km and longitudinally some 6000 km. It is separated into several discontinuous lobes (Fig.~\ref{fig01new}). The lobe that occupies the central part of the Medusae Fossae Formation is known as Lucus Planum \citep[e.g.][]{2011Icar..216..212K} (or alternatively, lobe B \citep{harrison2010mapping}. In the recently revised global geologic map of Mars \citep{tanaka2014geologic} two units make up Lucus Planum, namely the Hesperian and Amazonian-Hesperian transitional units (respectively {\it Htu} and {\it AHtu}) (Fig.~\ref{fig01new}) \citep{tanaka2014geologic}.

The MFF has been variously hypothesized to consist of pyroclastic flows \citep{1982JGR....87.1179S,2008JGRE..11312011M,2002JGRE..107.5058B}, pyroclastic airfall \citep{2000Icar..144..254T,2003JGRE..108.5111H,2011Icar..216..212K}, paleopolar deposits \citep{1988Icar...73...91S}, or atmospherically-deposited icy dust driven by climate cycles \citep{2004LPI....35.1635H}. A branching positive relief system within Lucus Planum was interpreted by \citet{2013P&SS...85..142H} as an ancient fluvial system originating from seepage sapping, implying that Lucus Planum was volatile-rich. The MFF shows evidence of a complex history of deposition, erosion and exhumation of both landforms and deposits \citep[e.g.][]{kerber2012progression}. Both erosional and depositional landforms are visible at different stratigraphic levels, resulting in complex morphologies.

Two sounding radars have been flown on Martian missions: MARSIS \citep{2005Sci...310.1925P} and SHARAD \citep{2007JGRE..112.5S05S}. Both instruments are synthetic aperture, low frequency radars carried by ESA's Mars Express and NASA's Mars Reconnaissance Orbiter, respectively. They transmit low-frequency radar pulses that penetrate below the surface, and are reflected by dielectric discontinuities in the subsurface. MARSIS is optimized for deep penetration, with a free-space range resolution of approximately 150 m, a footprint size of 10-20 km across-track and 5-10 km along-track. SHARAD has tenfold better resolution, at the cost of reduced penetration. Parts of the MFF have been probed by both of these sounding radars \citep{2007Sci...318.1125W,2009Icar..199..295C}, revealing a dielectric permittivity of the MFF material that is consistent with either a substantial component of water ice or a low-density, ice-poor material. While the work by \citet{2007Sci...318.1125W} was focused on Lucus Planum, estimates of dielectric properties by \citet{2009Icar..199..295C} were based on observations over Zephyria Planum, in the westernmost part of the Medusae Fossae Formation, and the area between Gordii Dorsum and Amazonis Mensa, at the Eastern end of the MFF.

The dielectric permittivity of the MFF material \citep{2007Sci...318.1125W,2009Icar..199..295C} is consistent with either a substantial component of water ice or a low-density, ice-poor material. There is no evidence for internal layering from SHARAD data \citep{2009Icar..199..295C}, despite the fact that layering at scales of tens of meters has been reported in many parts of the MFF \citep{2014LPI....45.2672K}. This lack of detection can be the result of one or more factors, such as high interface roughness, low dielectric contrast between materials, or discontinuity of the layers.

\section{Method}

\begin{figure}
\includegraphics[width=0.9\linewidth]{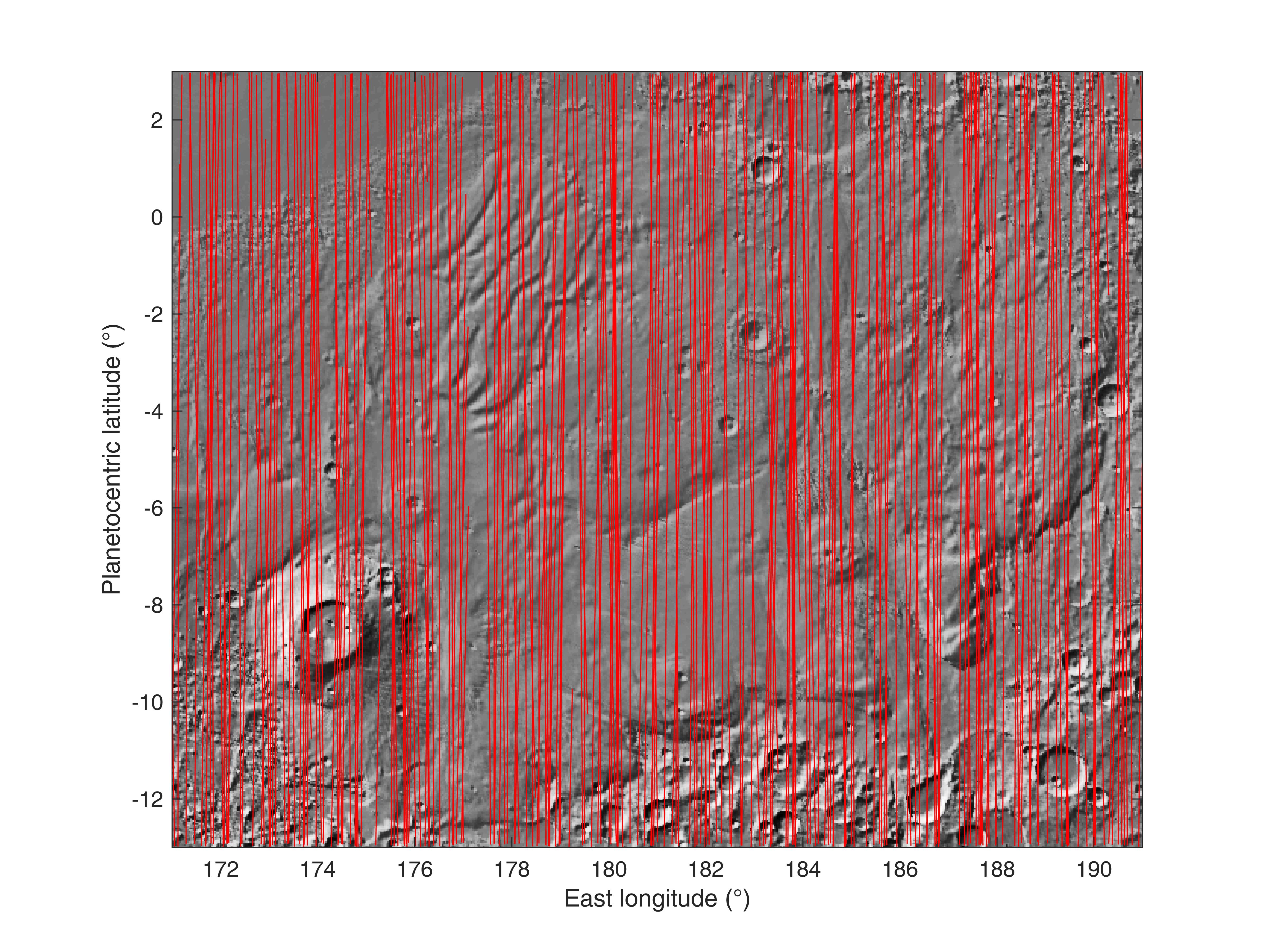}
\caption{Shaded relief map of the Lucus Planum area showing MARSIS ground tracks as black lines. Lines are thinner than the MARSIS swath width for legibility.}
\label{fig02new2}
\end{figure}

Operating since mid-2005, MARSIS has acquired 238 swaths of echoes across Lucus Planum, shown in Fig.~\ref{fig02new2}. Each swath consists of a few hundred observations, for a total of over 38,000 echoes. Data are affected by the dispersion and attenuation of the radar signal caused by ionospheric plasma, but a number of methods has been developed over the years to attenuate or compensate such effects \citep{2000SPIE.4084..624P,2003RaSc...38.1090A,2007GeoRL..3423204S,2008P&SS...56..917M,2009P&SS...57..393Z,2013SoSyR..47..430S,2016JGRE..121..180C}. Data used in this work have been processed using the methodology described by \citet{2013Icar..223..423C}, which consists in the maximization of the signal power in an interval centred around the strongest echo through the differential variation of the phase of the components of the Fourier signal spectrum.

\begin{figure}
\includegraphics[width=0.9\linewidth]{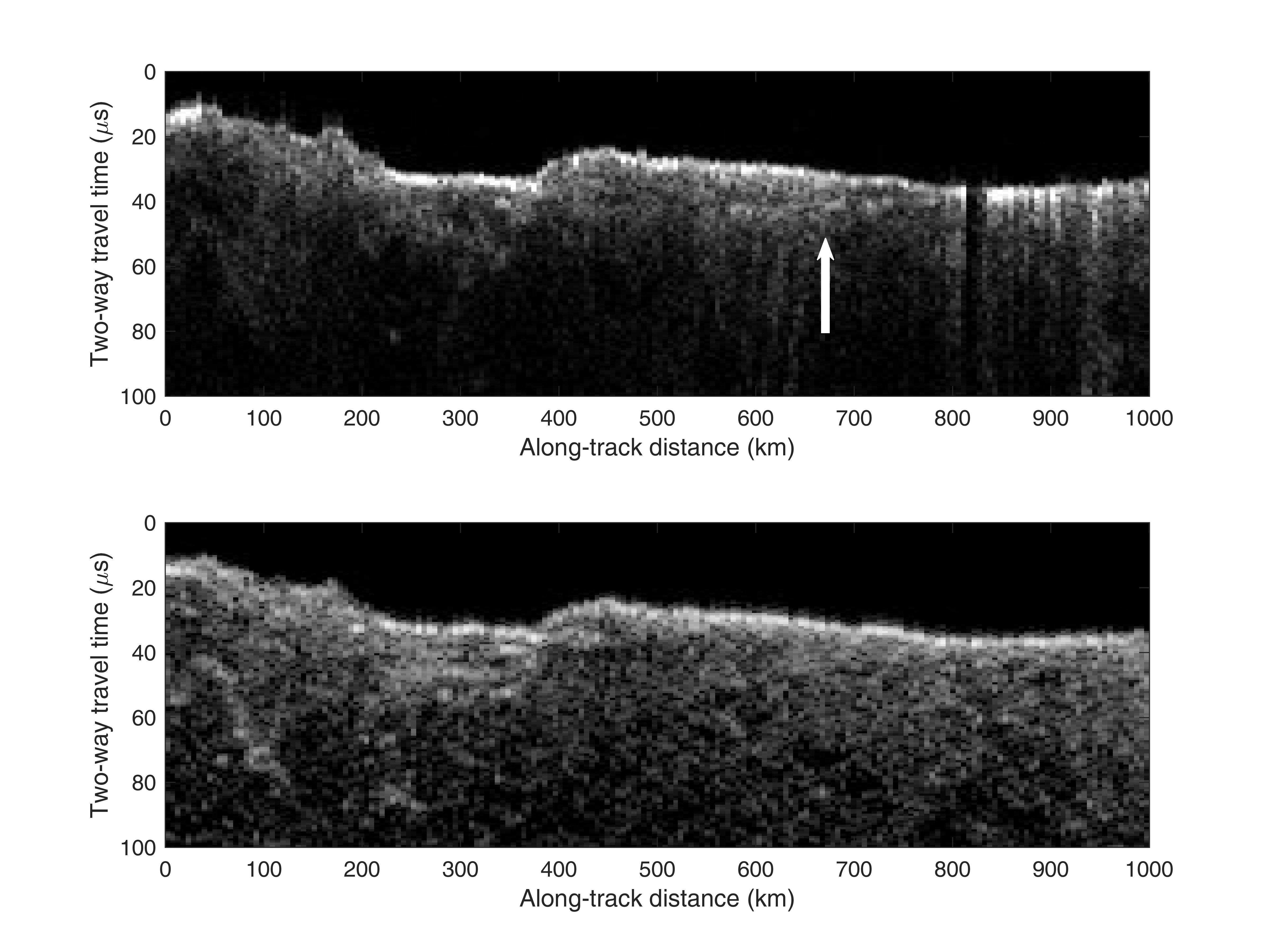}
\caption{Comparison between real (top) and simulated (bottom) radargrams for orbit 4011. The simulation reproduces echoes from topography only, while real data contain both surface and subsurface echoes. The arrow highlights a weak subsurface reflection.}
\label{fig03new}
\end{figure}

MARSIS data acquired continuously during the movement of the spacecraft are usually displayed in the form of radargrams, grey-scale images in which the horizontal dimension is distance along the ground track, the vertical one is the round trip time of the echo, and the brightness of the pixel is a function of the strength of the echo (ref. to example in Fig.~\ref{fig03new}). The first step in data analysis consisted in the visual inspection of radargrams to determine their quality. Observations were discarded if the ionospheric distortion compensation algorithm had failed, if spurious signals from the electronics of the spacecraft were present, or if exceptional ionosphere conditions resulted in a severe attenuation or absence of the signal. This reduced the number of radargrams suitable for further analysis by approximately 25\%.

The next step consisted in the identification of subsurface echoes in radargrams, which is complicated by the so called ``clutter'', that is by echoes coming from off-nadir surface features, such as craters or mountains, and reaching the radar after the nadir surface echo. As clutter can dwarf subsurface echoes, numerical electromagnetic models of surface scattering have been developed \citep[see e.g.][]{2004RaSc...39.1013N,2011P&SS...59.1222S} to validate the detection of subsurface interfaces in MARSIS data. They are used to produce simulations of surface echoes, which are then compared to the ones detected by the radar: any secondary echo visible in radargrams but not in simulations is interpreted as caused by subsurface reflectors (Fig.~\ref{fig03new}).

To analyse clutter, a code for the simulation of radar wave surface scattering was developed, based on the algorithm of \citet{2004RaSc...39.1013N}. The MOLA topographic dataset \citep{2001JGR...10623689S} was used to represent the Martian surface as a collection of flat plates called facets. Radar echoes were computed as the coherent sum of reflections from all facets illuminated by the radar. The computational burden of simulations required the use of the SuperMUC supercomputer at the Leibniz--Rechenzentrum, Garching, Germany.

Subsurface reflections in Lucus Planum are usually weak and often have a diffuse appearance (Fig.~\ref{fig03new}). Several methods were attempted to automatically identify such reflections in radargrams, but eventually a supervised procedure was used, in which an operator manually selects a few points marking the position of the interface in a radargram, and then the procedure itself outlines the interface and records its aerocentric coordinates, its time delay from the surface echo, and its reflected power. The confidence in the retrieved coordinates is based on the accuracy of the reconstructed Mars Express trajectory, which is estimated to be a fraction of the MARSIS footprint size. To better determine the position and power of subsurface echoes, radar signals have been interpolated with the Fourier interpolation method to reduce the sampling interval to 0.1 $\mu$s. The precision in the determination of the time delay is assumed to be the one-way delay resolution (or 0.5 $\mu$s, corresponding to 150 m free-space), while the uncertainty in echo power is considered to be below 0.5 dB because of the interpolation.

\section{Results}

\begin{figure}
\includegraphics[width=0.9\linewidth]{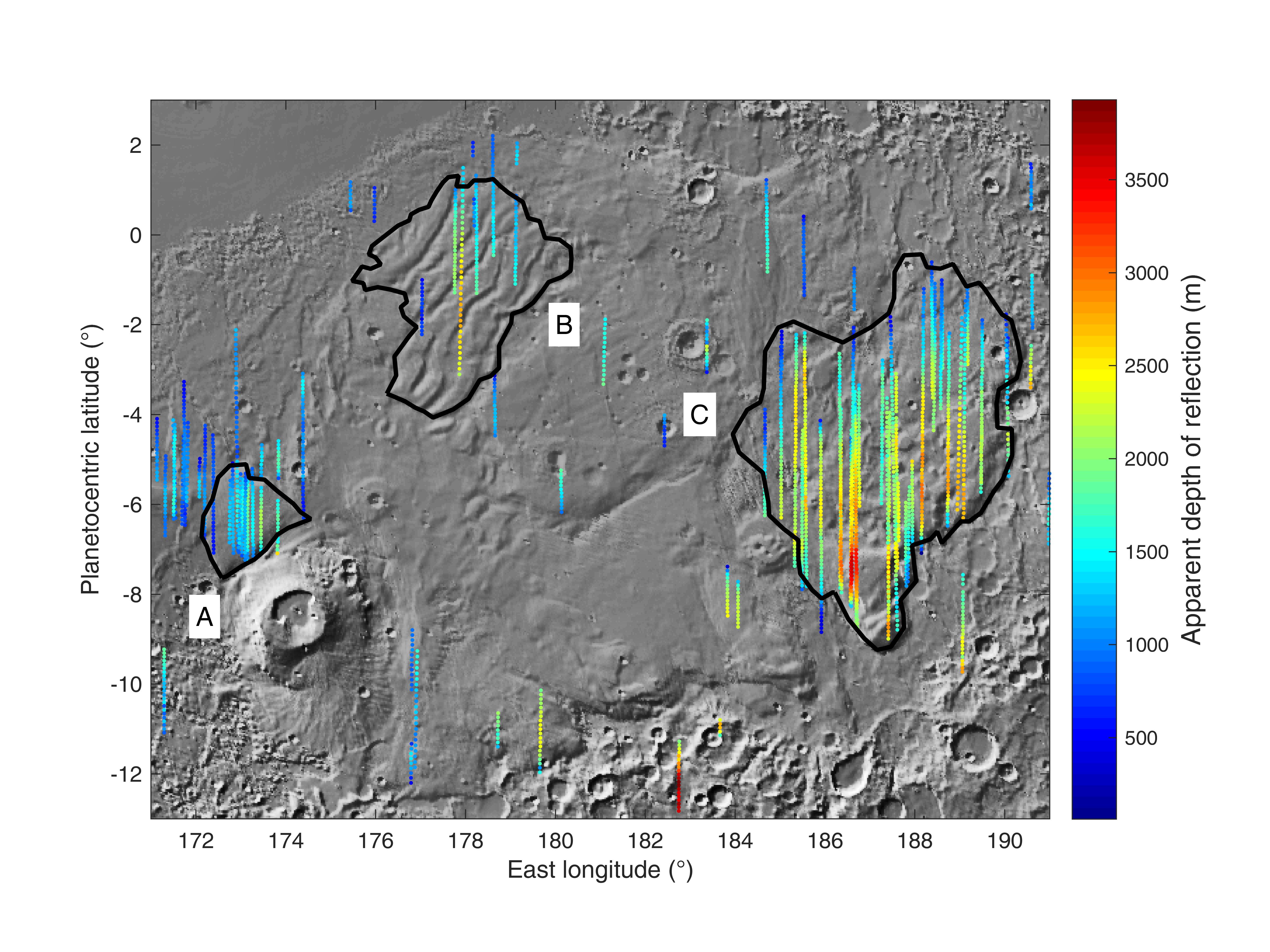}
\caption{Map of the location and apparent depth of subsurface echoes in Lucus Planum. Apparent depth is coded according to the colour scale on the right, and differs from real depth by a factor $\sqrt{\varepsilon}$ (see Eq.~\ref{delay2depth}). Black contours outline what appear to be different morphologic provinces within Lucus Planum.}
\label{fig04new}
\end{figure}

A total of 97 subsurface reflectors were identified, extending along track over distances up to 500 kilometres. Their distribution across Lucus Planum is shown in Fig.~\ref{fig04new}. In spite of several high-quality radargrams crossing the central part of Lucus Planum, only a handful of subsurface interfaces could be detected there, most of which are shallow, often associated with pedestal craters. Reflectors concentrate in specific areas: the deposits on the north-western flanks of Apollinaris Patera, the rugged terrain North of Tartarus Scopulus and the large lobe located North-East of Memnonia Sulci. The contours of these areas follow closely morphologically distinct provinces within Lucus Planum, which suggests that variations in surface morphology could be tied to changes in the material forming Lucus Planum. These areas are outlined in Fig.~\ref{fig04new} and labelled ``A'', ``B'' and ``C'', respectively.

Figure~\ref{fig04new} shows the apparent depth of reflectors, estimated from the measured round-trip time delay between surface and subsurface echo by:

\begin{equation}
\label{delay2depth}
z = \frac{c \; \tau}{ 2 \sqrt{ \varepsilon } }
\end{equation}

\noindent
where $z$ is depth, $c$ the speed of light \textit{in vacuo}, $\tau$ the round-trip time delay between surface and subsurface echo, and $\varepsilon$ is the real part of the relative complex permittivity (also called dielectric constant) of the Lucus Planum material. The apparent depth $z_a$ was computed assuming that $\varepsilon$ is equal to 1, corresponding to the permittivity of free space:

\begin{equation}
\label{zapp}
z_a = \frac{c \; \tau}{ 2 }
\end{equation}

Apparent depths overestimate the thickness of Lucus Planum by a factor comprised between $\sqrt{3}$ and 3, depending on the nature of the material through which the wave propagates \citep[][Appendix E]{ulaby1986microwave}.

\begin{figure}
\includegraphics[width=0.9\linewidth]{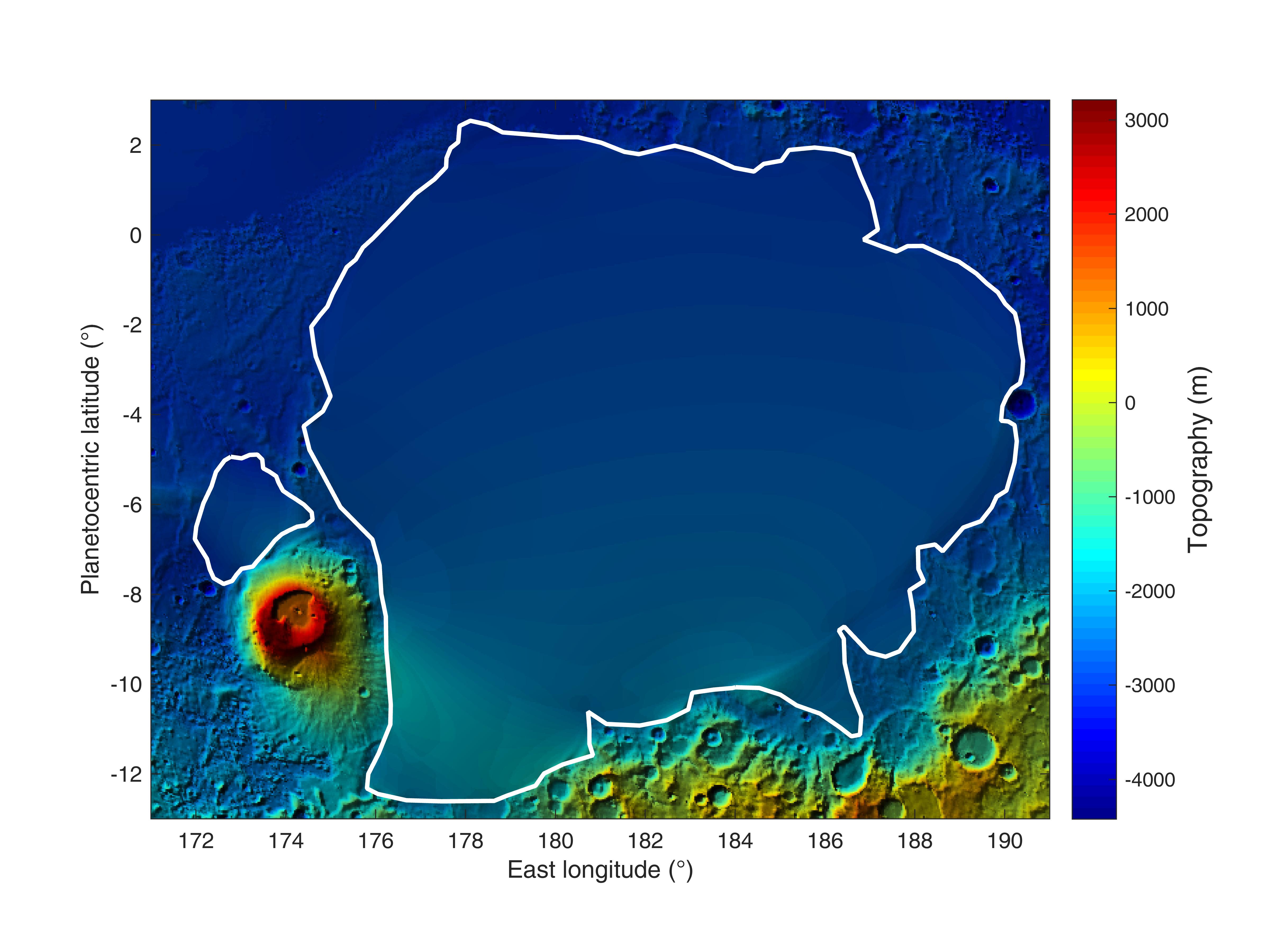}
\caption{Topographic map of the Lucus Planum area, based on the MOLA dataset. The white contours identify the areas in which MOLA topography was removed, and then the base of the Lucus Planum and Apollinaris Patera deposits was interpolated from the remaining topographic information through the natural neighbour method.}
\label{fig05new}
\end{figure}

Estimates of permittivity for the different regions of Lucus Planum provide some insight on their nature and a more precise evaluation of their thickness. Following the approach first presented in \citet{2005Sci...310.1925P} and used also in \citet{2007Sci...318.1125W}, we produced an independent estimate of the thickness of Lucus Planum assuming that the deposits rest on a surface in lateral continuity with the surrounding topography, and that MARSIS echoes come from such surface. The white contours in Fig.~\ref{fig05new} encompass those areas in which MOLA topography was removed, and then interpolated from the remaining topographic information through the natural neighbour method \citep{natural}.

\begin{figure}
\includegraphics[width=0.4\linewidth]{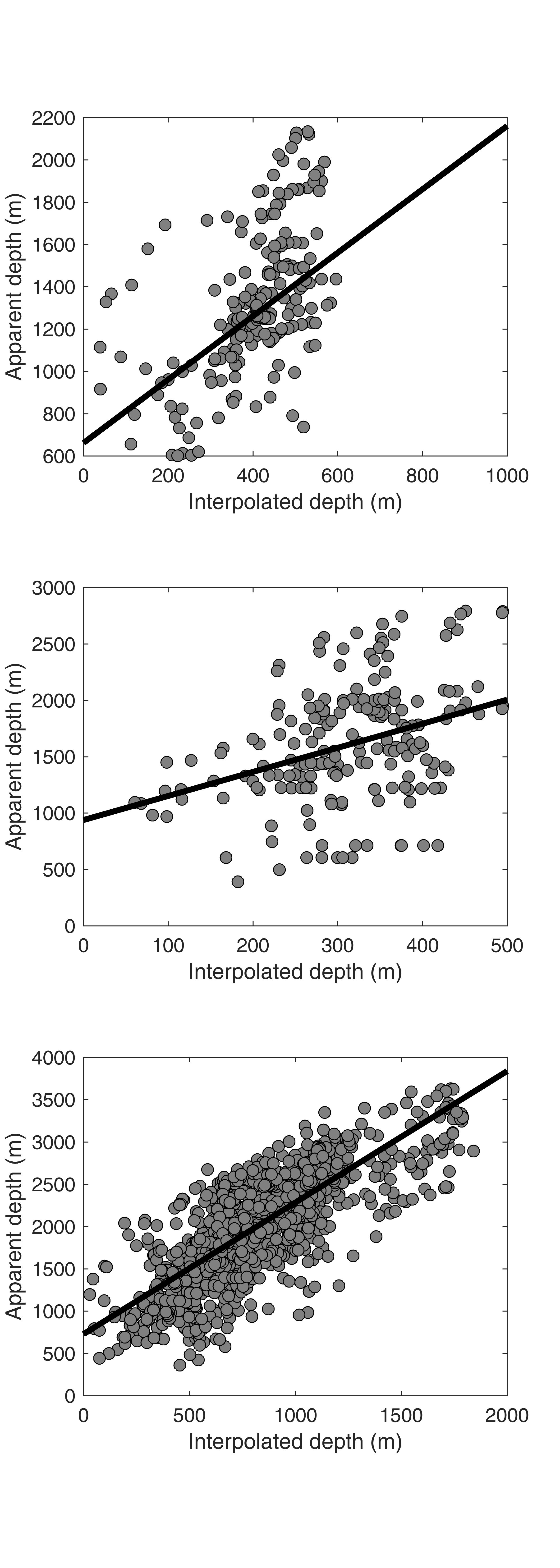}
\caption{Plot of the thickness of Lucus Planum, estimated by interpolating the surrounding topography, versus the apparent thickness derived from the delay time of subsurface radar echoes. Graphs refer, from top to bottom, to areas A, B and C (see Fig.~\ref{fig04new}). The best-fit line has been computed using the least absolute deviations method.}
\label{fig06new}
\end{figure}

The difference between the actual topography and the interpolated basal topography of Lucus Planum provides an estimate of the depth of the base of Lucus Planum, $z_i$. By inserting $z_i$ in Eq.~\ref{delay2depth}, solving Eq.~\ref{zapp} by $c\tau$, and rearranging and simplifying equal terms, we obtain:

\begin{equation}
\label{best_fit}
z_a = \sqrt{ \varepsilon } z_i
\end{equation}

\noindent
from which we see that the slope of the best-fit line in a plot of interpolated vs. apparent depth provides an estimate of $\sqrt{ \varepsilon }$. The resulting plots for areas A, B and C are shown in Fig.~\ref{fig06new}. Because of the large dispersion of data points in some of the plots, the best-fit line was computed using the least absolute deviations method \citep{LAD}, which is less sensitive to outliers than the least squares method.

The slopes and constant terms of the best-fit lines in Fig.~\ref{fig06new} are reported in Table~\ref{table1}. For each value, the corresponding 95\% confidence interval of the estimate is listed, providing some insight on the quality of the data fit. Table~\ref{table1} reports also estimates of $\varepsilon$, obtained from the values of slopes and their corresponding 95\% confidence bounds through Eq.~\ref{best_fit}.

\begin{table}
\caption{Coefficients of the best-fit lines in Fig.~\ref{fig06new}, together with their 95\% confidence bounds. The table reports also the corresponding estimates of $\varepsilon$ and of its confidence interval, derived from Eq.~\ref{best_fit}.}
\label{table1}
\begin{tabular}{ccc}
\hline
Area & Slope                                    & \parbox{2.5cm}{\centering 95\% confidence interval} \\
\hline
A    & 1.50                                     & [ 1.17, 1.82 ]                                      \\
B    & 2.13                                     & [ 1.34, 2.93 ]                                      \\
C    & 1.56                                     & [ 1.49, 1.62 ]                                      \\
\hline
Area & \parbox{1.5cm}{\centering Constant term} & \parbox{2.5cm}{\centering 95\% confidence interval} \\
\hline
A    & 661.2                                    & [ 527.9,  794.5 ]                                   \\
B    & 937.6                                    & [ 679.1, 1196.1 ]                                   \\
C    & 727.1                                    & [ 671.4,  782.8 ]                                   \\
\hline
Area & $\varepsilon$                            & \parbox{2.5cm}{\centering 95\% confidence interval} \\
\hline
A    & 2.25                                     & [ 1.38, 3.33 ]                                      \\
B    & 4.56                                     & [ 1.80, 8.57 ]                                      \\
C    & 2.42                                     & [ 2.23, 2.62 ]                                      \\
\hline
\end{tabular}
\end{table}

From Eq.~\ref{best_fit}, the value of the constant term in best-fit lines should be zero, different from what is reported in Table~\ref{table1}. The presence of a constant term indicates a systematic error in the evaluation of $z_a$, $z_i$ or both. Because the range resolution of MARSIS is about 150 m in free space \citep{2005Sci...310.1925P}, the constant terms in Table~\ref{table1} correspond to a few range resolution cells. One possible explanation is that the interpolation method failed to provide a correct estimate of the basal topography: because Lucus Planum straddles the dichotomy boundary, the topography beneath it is expected to be complex, affecting the precision of results. Another possibility is a systematic overestimation of the time delay of subsurface echoes in radargrams, perhaps because subsurface reflections are less sharp than surface ones, and the manual determination of their exact position introduces additional uncertainties.

Permittivity is a complex quantity: its real part affects the velocity of an electromagnetic wave, while its imaginary part is related to the dissipation (or loss) of energy within the medium. The ratio between the imaginary and the real part of the complex permittivity is called the loss tangent. Estimating the loss tangent of the material within Lucus Planum provides an additional constraint on its nature and can be used by way of checking on the significance of the values of $\varepsilon$ in Table~\ref{table1}.

The loss tangent over parts of the Medusae Fossae formation was estimated from the rate of decay of the subsurface echo power as a function of time delay by \citet{2007Sci...318.1125W}. Following a similar approach, we assumed that the surface and the subsurface interfaces over Lucus Planum are smooth at MARSIS frequencies, meaning that the RMS height of topography is a fraction of the wavelength, and that Lucus Planum consists only of non-magnetic, low loss material. While a higher roughness would cause only a fluctuation of surface and subsurface power without affecting the mean rate of subsurface power decay with depth, the assumption that Lucus Planum consists of a low loss, non-magnetic material is validated by previous results \citep{2005Sci...310.1925P}, and would result in little or no subsurface interface detections if violated. Under these assumptions, following \citet{1974IEEEP..62..769P}, the surface echo power $P_s$ can be written as follows:

\begin{equation}
\label{Psurf}
P_s = P_t \cdot \left( \frac{ G \lambda }{ 8 \pi H } \right)^2 \cdot \left| R_s \right|^2
\end{equation}

\noindent
with $P_t$ the transmitted power, $G$ the antenna gain, $\lambda$ the wavelength, $H$ the spacecraft altitude and $R_s$ the surface Fresnel reflection coefficient at normal incidence. Analogously, the subsurface echo power $P_{ss}$ can be computed through the following expression:

\begin{eqnarray}
\nonumber
P_{ss} & = & P_t \cdot \left( \frac{ G \lambda }{ 8 \pi ( H + z ) } \right)^2 \cdot \left( 1 - \left| R_s \right|^2 \right)^2 \cdot \\
\label{Psubsurf}
       &   & \left| R_{ss} \right|^2 \cdot \exp \left( -2 \pi f \tan\delta \tau \right)
\end{eqnarray}

\noindent
where $R_{ss}$ is the subsurface Fresnel reflection coefficient at normal incidence, $f$ the radar frequency, $\tan\delta$ the loss tangent of the Lucus Planum material, here assumed to be constant through its entire thickness, while $z$ and $\tau$ have been defined in Eq.~\ref{delay2depth}.

By dividing Eq.~\ref{Psubsurf} and Eq.~\ref{Psurf}, and then taking the natural logarithm of the result, the following expression is obtained:

\begin{equation}
\label{lnPssPs}
\ln \left( \frac{ P_{ss} }{ P_s } \right) = -2 \pi f \tan\delta \tau + K
\end{equation}

\noindent
where $K$ is a term depending on $R_s$ and $R_{ss}$. The topography of Lucus Planum is characterized by a roughness that is not negligible compared to the MARSIS wavelength \citep[see][]{2000JGR...10526695K,2003GeoRL..30.1561N}. This implies that $P_s$ and $P_{ss}$ fluctuate around a mean value that is a function of statistical parameters characterizing the topography, such as RMS height and RMS slope \citep[see for example][]{ogilvy}. Under the assumption that such parameters do not vary significantly within each of the three areas A, B and C, then roughness will cause only a variation of the value of parameter $K$ and the addition of a random noise to $\ln \left( P_{ss} / P_s \right)$ in Eq.~\ref{lnPssPs}.

Other factors connected to the internal structure of the Lucus Planum and Apollinaris Patera deposits are unlikely to affect Eq.~\ref{lnPssPs} significantly. A surface layer thinner than the vertical resolution of the radar can generate interferences so as to drastically reduce surface reflectivity, as in the case of the CO$_2$ layer over the SPLD identified by \citet{2009Icar..201..454M}. However, such coherent effects require a very smooth surface and are strongly frequency-dependent. Both the rougher surface of Lucus Planum and Apollinaris Patera \citep{2000JGR...10526695K,2003GeoRL..30.1561N}, and the fact that such dependence on frequency was not found in the data seem to rule out the presence of such a layer.

Other material inhomogeneities in the dielectric properties at depths below the vertical resolution of MARSIS would tend to produce surface echoes whose power is dominated by the dielectric permittivity of the layers closest to the surface, as discussed in \citet{2014GeoRL..41.6787G}. This effect would alter $P_{ss}/P_s$, but it would not change the rate at which this quantity decreases with depth, that is the first term of the right side of Eq.~\ref{lnPssPs}. Random inhomogeneities within the deposits, whose characteristic size is comparable to the MARSIS wavelength, would result in volume scattering, that is in the diffusion of electromagnetic radiation within the deposits away from the direction of propagation.

The diffuse, weak echoes between surface and basal reflections visible in Fig.~\ref{fig02new2} could be caused by volume scattering, although they could also originate from surface roughness. Volume scattering cannot be easily characterized from the measure of backscattered radiation, but it would attenuate the subsurface radar echo. This effect cannot be separated from dielectric attenuation, and it would thus lead to a systematic overestimate of $\tan\delta$ from Eq.~\ref{lnPssPs}, which thus constitutes an upper bound for the true dielectric attenuation.

\begin{figure}
\includegraphics[width=0.4\linewidth]{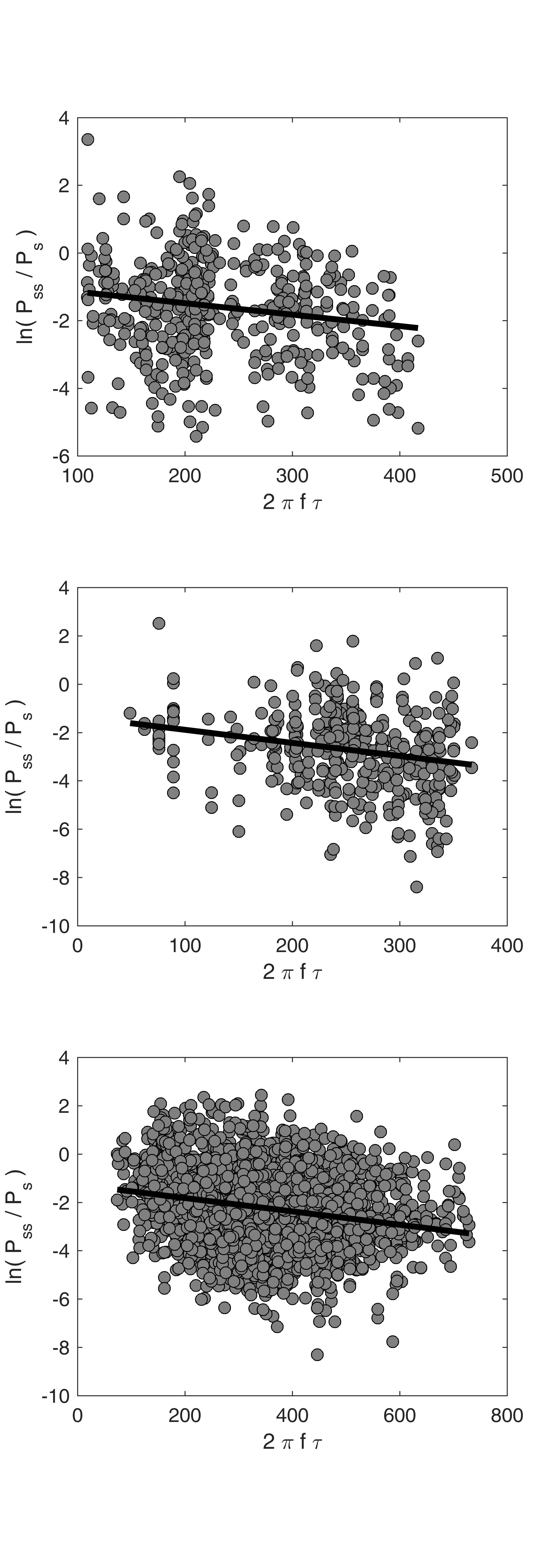}
\caption{Plot of the number of cycles completed by the radar wave within Lucus Planum versus the natural logarithm of the subsurface to surface echo power ratio. Graphs refer, from top to bottom, to areas A, B and C (see Fig.~\ref{fig04new}). The best-fit line has been computed using the least absolute deviations method.}
\label{fig07new}
\end{figure}

With these caveats, the slope of the best-fit line in a plot of $2 \pi f \tau$ (that is the number of cycles completed by the radar wave within Lucus Planum) vs. the natural logarithm of the subsurface to surface echo power ratio will provide an estimate of $\tan\delta$. Such plots for areas A, B and C, with the corresponding best-fit lines, are shown in Fig.~\ref{fig07new}. In analogy with the estimation of $\varepsilon$, the best-fit line was computed using the least absolute deviations method.

The slopes and constant terms of the best-fit lines in Fig.~\ref{fig07new} are reported in Table~\ref{table2}. For each value we report the corresponding 95\% confidence interval of the estimate, to provide some insight on the quality of the data fit. Table~\ref{table2} reports also the corresponding estimates of $\tan\delta$ with their 95\% confidence bounds.

\begin{table}
\caption{Coefficients of the best-fit lines in Fig.~\ref{fig07new}, together with their 95\% confidence bounds. The table reports also the corresponding estimates of $\tan\delta$ and of its confidence interval, derived from Eq.~\ref{lnPssPs}.}
\label{table2}
\begin{tabular}{ccc}
\hline
Area & Slope                                    & \parbox{2.5cm}{\centering 95\% confidence interval} \\
\hline
A    & -0.0034                                  & [ -0.0052, -0.0016 ]                                \\
B    & -0.0054                                  & [ -0.0077, -0.0031 ]                                \\
C    & -0.0028                                  & [ -0.0032, -0.0023 ]                                \\
\hline
Area & \parbox{1.5cm}{\centering Constant term} & \parbox{2.5cm}{\centering 95\% confidence interval} \\
\hline
A    & -0.809                                   & [ -1.255, -0.363 ]                                  \\
B    & -1.349                                   & [ -1.933, -0.764 ]                                  \\
C    & -1.270                                   & [ -1.436, -1.104 ]                                  \\
\hline
Area & $\tan\delta$                             & \parbox{2.5cm}{\centering 95\% confidence interval} \\
\hline
A    & 0.0034                                   & [ 0.0016, 0.0052 ]                                  \\
B    & 0.0054                                   & [ 0.0031, 0.0077 ]                                  \\
C    & 0.0028                                   & [ 0.0023, 0.0032 ]                                  \\
\hline
\end{tabular}
\end{table}

\section{Discussion}

The lack of subsurface reflections in the central part of Lucus Planum can be the result of several factors, some of which depend on surface properties. A high topographic roughness at scales comparable to the radar wavelength causes scattering of the incident pulse, resulting in weaker surface and subsurface echoes. However, RMS heights estimated from MOLA data both over baselines of a few to several kilometers \citep{2000JGR...10526695K} and within the MOLA footprint \citep{2003GeoRL..30.1561N} are higher in area C, where subsurface detections are frequent, than in the central part of Lucus Planum. Another possibility is that the basal roughness is higher in its central part. Because subsurface echoes appear to be associated with areas of distinct surface morphology, a third possibility is that the central part of Lucus Planum consists of denser, more radar-attenuating material.

Values of $\varepsilon$ in areas A and C are similar to those found by \citet{2007Sci...318.1125W} and \citet{2009Icar..199..295C}, while those in area B appears to be higher, although the estimate is affected by a larger uncertainty. The same trend, both in values and confidence intervals, is observed also for $\tan \delta$. It can also be seen in Fig.~\ref{fig04new} that the spatial density of subsurface interface detections is much higher in areas A and C than in area B, in spite of a comparable density of coverage (see Fig.~\ref{fig02new2}). Surface roughness at kilometre scale in area B is similar to that of the central part of Lucus Planum and smaller than that of area C \citep{2000JGR...10526695K}, in spite of the different surface morphology, while roughness in area B at hundred-meters scale is comparable to that of area C \citep{2003GeoRL..30.1561N}. Because the dearth and weakness of subsurface echoes in area B do not correlate with a higher surface roughness compared to areas A and C, we favour the interpretation that, in spite of the large uncertainties, the higher value of the complex relative permittivity in area B is an indication of a change in bulk dielectric properties with respect to areas A and C.

The relative dielectric constant of volcanic rocks such as those thought to constitute the Martian crust is variable, ranging between 2.5 for pumice to about 10 or even higher for dense basalts \citep[][Appendix E]{ulaby1986microwave}. Following a search in the literature and a set of new measurements between 0.01 and 10 MHz, \citet{1999JVGR...91...79R} concluded that the main factor in determining the value of $\varepsilon$ is porosity, finding the following empirical relation for dacitic rocks:

\begin{equation}
\label{epsrock}
\left( \varepsilon \right)^{0.96} = \Phi + 6.51 \left( 1 - \Phi \right)
\end{equation}

\noindent
where $\Phi$ is porosity. Such relation is similar to estimates for other non-basaltic rocks, and holds also for basalts, although with a greater variability due perhaps to Fe-Ti oxide mineral content \citep{1999JVGR...91...79R}. Modelling the dependence of $\varepsilon$ on porosity through Eq.~\ref{epsrock}, and inverting such equation to obtain estimates of $\Phi$, it is found that the values of $\varepsilon$ in Table~\ref{table1} are consistent with a porosity between 0.6 and 0.9 for area A, up to 0.85 for area B, and between 0.7 and 0.8 for area C. Such high values are typical of volcanic rocks extruded through explosive, rather than effusive, processes. In their study of the correlation between the distribution of porosity in pyroclasts and eruption styles, \citet{2011JVGR..203..168M} found that porosity values above 0.5 are characteristic of the product of explosive basaltic eruptions \citep[eruptions of gas-rich, low-viscosity magma, e.g.][]{wilson1994mars} or even highly explosive subplinian, plinian or ultraplinian eruptions, whose deposits derive from fall--out or pyroclastic density currents.

Potential sources of Martian pyroclastic deposits have been discussed in the literature \citep[e.g.][]{2003JGRE..108.5111H,2009Icar..199..295C} and might be largely related to Tharsis. Also Apollinaris Patera is at close reach for Lucus Planum \citep{2011Icar..216..212K}. The role of possibly buried volcanic edifices has been suggested by \citet{2009Icar..199..295C}: evidence of such edifices has not been found so far in the area.

Tridymite has recently been discovered by the Curiosity rover within lacustrine sediments in Gale Crater \citep{morris2016pnas}, suggesting the presence of silica--rich volcanics within the crater's watershed. If the Medusae Fossae Formation is composed of explosive volcanic material, the upper portion of Mount Sharp, which is thought to be a part of the MFF \citep{2011Icar..214..413T,tanaka2014geologic}, could be a source for that material.

Another plausible explanation of the nature of Lucus Planum materials and the Medusae Fossae Formation in general is that they might consist of ice--rich dust or ice--laden porous rock, although previous estimates of dielectric properties based on radar data proved inconclusive \citep{2007Sci...318.1125W,2009Icar..199..295C}.

The permittivity of a mixture of ice and dust can be estimated using a mixing formula. Because of the lack of knowledge about the size and shape of pores or ice inclusions in the rock, in the following analysis we selected the general Polder--van Santen model \citep{polvan46}. This formula is one of the simplest and yet more widely used, and it has the special property that it treats the inclusions and the hosting material symmetrically; it balances both mixing components with respect to the unknown effective medium, using the volume fraction of each component as a weight:

\begin{equation}
\label{polvanmix}
( 1 - f ) \; \frac{ \varepsilon_{\text{h}} - \varepsilon_{\text{eff}} }{ \varepsilon_{\text{h}} + 2 \varepsilon_{\text{eff}} } + f \; \frac{ \varepsilon_{\text{i}} - \varepsilon_{\text{eff}} }{ \varepsilon_{\text{i}} + 2 \varepsilon_{\text{eff}} } = 0
\end{equation}

\noindent
where $f$ is the volume fraction of inclusions in the mixture, $\varepsilon_{\text{h}}$ is the permittivity of the host material, $\varepsilon_{\text{i}}$ that of the inclusions, and $\varepsilon_{\text{eff}}$ the effective permittivity of the mixture.

Water ice has a relative dielectric constant well within the range of values typical of porous rocks (2--6), while its loss tangent can vary by orders of magnitude as a function of temperature in the range 100-270 K, which is applicable to Martian conditions. Using the empirical formulas presented in \citet{1998ASSL..227..241M} and a mean surface temperature of 210 K, typical for the latitudes of Lucus Planum according to \citet{2004Icar..169..324M}, it is found that the real part of the permittivity of water ice is $\approx 3.1$, and the loss tangent is $\approx 5 \cdot 10^{-5}$. We hypothesized that the relative dielectric constant of the rocky component in the Lucus Planum material could range from 7 to 15 \citep{1999JVGR...91...79R}, and that its loss tangent could independently vary between $10^{-3}$ and $10^{-1}$ \citep[][Appendix E]{ulaby1986microwave}. The Polder--van Santen mixing rule was then used to model the effective permittivity of all possible combinations of relative dielectric constant, loss tangent and porosity, similarly to the method described in \citet{2012JGRE..117.9008A}.

Comparing the results with the estimated values in areas A, B and C, we found that no mixture of rock and ice could produce a complex permittivity compatible with that of areas A and C. It is possible to obtain compatible permittivity values for these two areas using a three-component mixture, that is rock, ice and void, but the significance of this result, given the weakly constrained multi-dimensional parameter space, is difficult to assess. In the map of water-equivalent hydrogen content for the Martian soil produced by \citet{2004JGRE..109.9006F}, the Lucus Planum area appears to be relatively water--rich, with a water--equivalent hydrogen content estimated at around 8\%. This value however is referred to the first meter of depth, while the dielectric permittivity derived from MARSIS data is an average over the whole thickness of the Lucus Planum and Apollinaris Patera deposits.

For area B, mixtures with an ice volume fraction between 0.3 and 0.9 and a loss tangent for the rocky material comprised between $3 \cdot 10^{-3}$ and $3 \cdot 10^{-2}$ could return a range of permittivity values consistent with estimates. To determine the significance of this result, we also computed the effective permittivity of a mixture of rock and void (empty pores) over the same parameter space. We found that values consistent with those of area B could be obtained for a range of porosity and loss tangent values similar to that of the mixture of rock and ice. We thus conclude that the nature of the bulk material in area B cannot be reliably determined using only the data provided by this analysis.

\begin{figure}
\includegraphics[width=0.9\linewidth]{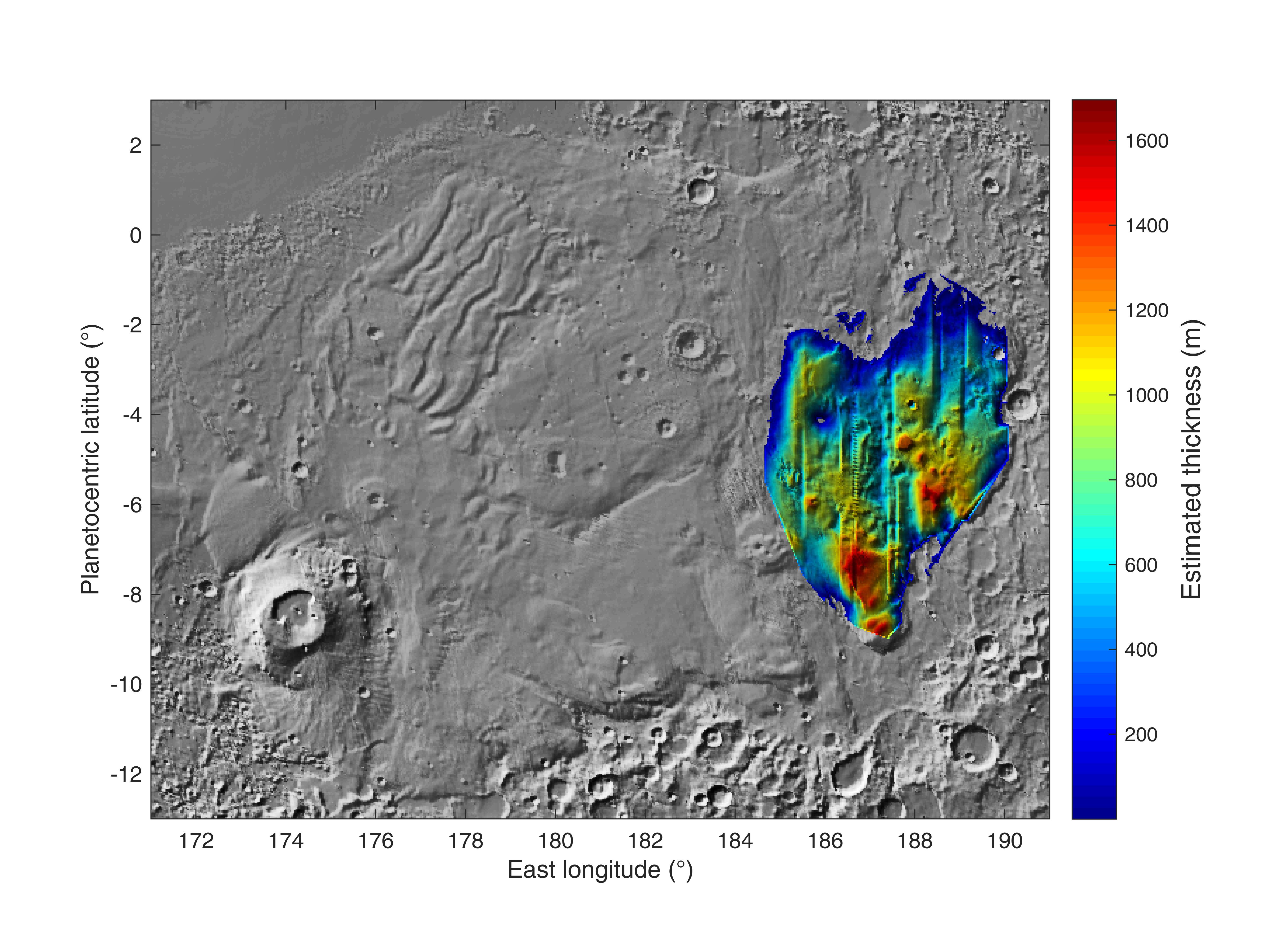}
\caption{Map of the thickness of the Eastern part of Lucus Planum, superimposed to a shaded relief map covering the same area shown in Fig.~\ref{fig04new}. Thickness has been determined using the estimated relative dielectric constant in Table~\ref{table1}, and interpolated through the natural neighbour method.}
\label{fig08new}
\end{figure}

Area C presents the highest number and density of subsurface detections, and the smallest uncertainty in the estimates of dielectric properties. We therefore inserted in Eq.~\ref{delay2depth} the value of $\varepsilon$ from Table~\ref{table1} to estimate the thickness of the Lucus Planum deposits in such area, and then interpolated this quantity over area C through the natural neighbour method. The result is included in Fig.~\ref{fig08new}, in which the colour--coded thickness is layered on a shaded relief map of Lucus Planum. The deposits are several hundred meters thick on average, locally reaching a thickness up to 1.5 kilometres, for a total volume of $\approx 6.8 \cdot 10^4$ km$^3$. The deposit thickness varies positively with regional elevations, being higher in the South, and lower in the North.

Overall, Lucus Planum subsurface as sounded by MARSIS appears to be locally to regionally inhomogeneous. This can be interpreted in terms of complex, multi--process components of the deposits constituting Lucus Planum and possibly the Medusa Fossae Formation as a whole \citep[e.g.][]{kerber2010icarus,2014LPI....45.2672K}. An interplay between a possibly dominating volcano--sedimentary component and local, possibly late--stage erosional and partially depositional episodes could be envisaged. Such episodes on or in the vicinity of Lucus Planum likely occurred in the relatively recent past \citep{2013P&SS...85..142H}, leading to extensive resedimentation of Lucus Planum materials \citep{kerber2012progression}.

The dielectric properties of the north--western part of Lucus Planum, implying a higher density compared to the other radar--transparent areas, and the inferred strong attenuation of the radar signal in its central part could be interpreted as due to the presence of indurated sedimentary deposits. Their existence within Lucus Planum is consistent with extensive reworking of those deposits through time \citep{2013P&SS...85..142H,kerber2012progression,2014LPI....45.2672K}. Although such deposits could be compositionally similar to the overall MFF materials \citep{2013P&SS...85..142H}, they could be locally remobilised, thus changing their architecture, structure and texture, including their degree of cementation \citep{kerber2012progression}. Their Hesperian--Amazonian age \citep{kerber2010icarus,tanaka2014geologic} would match both late stage valley network as well as vigorous Tharsis activity. MARSIS data cannot shed much light on small-- to medium--scale lateral and vertical variations, which will require additional work at an appropriate scale.

The possibility of sampling with Mars Science Laboratory on its way uphill on Mount Sharp in Gale Crater some material, even not necessarily in--situ but made available through mass wasting and resedimentation, would allow for some indirect ground truth: pyroclastic, possibly acidic, volcanic material of an age comparable with that of Lucus Planum (Hesperian to Amazonian) would offer support to a pyroclastic origin of the MFF. On the other hand, resedimented material so far from the MFF main bodies would not allow for volatiles to be embedded and preserved.

\section{Conclusions}

MARSIS acquired 238 radar swaths across Lucus Planum, providing sufficient coverage for the study of the internal structure and dielectric properties of this part of the MFF. Subsurface reflections were found only in three areas, marked by a distinctive surface morphology, while the central part of Lucus Planum appears to be made of radar--attenuating material preventing the detection of basal echoes. The bulk dielectric constant of these areas was estimated by comparing their apparent thickness from radar data with their basal topography, extrapolated from the surrounding terrains. The complex part of the dielectric permittivity was derived from the weakening of basal echoes as a function of apparent depth, yielding results that are consistent with the estimated dielectric constant. The inferred bulk properties were compared with known materials such as volcanic rocks and ice--dust mixtures. The interpretation that the eastern area of Lucus Planum and the deposits on the north--western flanks of Apollinaris Patera consist of high--porosity pyroclastic material is strongly supported by results, while north--western Lucus Planum is likely to be much less porous. No conclusion could be drawn about the presence of pore ice.

All evidence points to Lucus Planum being highly inhomogeneous. The exact origin of the deposits cannot be constrained by radar data alone, but our results are consistent with an overall pyroclastic origin as suggested by \citet{2009Icar..199..295C} and \citet{2007Sci...318.1125W} for the MFF. The geological complexity of the subsurface revealed through MARSIS data is consistent with a combination of processes acting through space and time, including fluvial (possibly outflow--related) activity occurred during the emplacement of the {\it Htu} and {\it AHtu} units \citep{tanaka2014geologic}, as well as eolian deposition \citep{kerber2012progression}. The overall surface textural and topographical heterogeneity might be linked to post--emplacement erosional processes related to regional wind dynamics \citep{kerber2012progression} on a variably indurated substrate. The evidence in this work was not sufficient to demonstrate the presence of an ice-related component in the central part of Lucus Planum, although this cannot be conclusively excluded. A full understanding of such a complex geological history will require the integration of several datasets at different scales and with different resolutions.


\begin{acknowledgments}
This work was supported by the Italian Space Agency (ASI) through contract no. I/032/12/1. The numerical code for the simulation of surface scattering was developed at the Consorzio Interuniversitario per il Calcolo Automatico dell'Italia Nord--Orientale (CINECA) in Bologna, Italy. Simulations were produced thanks to the Partnership for Advanced Computing in Europe (PRACE), awarding us access to the SuperMUC computer at the Leibniz--Rechenzentrum, Garching, Germany through project 2013091832. Test simulations were run Jacobs University CLAMV HPC cluster, and we are grateful to Achim Gelessus for his support. This research has made use of NASA's Astrophysics Data System. APR has been supported by the European Union FP7 and Horizon 2020 research and innovation programmes under grant agreements \#654367 (EarthServer--2) and \#283610 (EarthServer). Data used in this analysis have been taken from the MARSIS public data archive, which is currently undergoing validation before publication on the Planetary Science Archive of the European Space Agency (\verb|http://www.cosmos.esa.int/web/psa/mars-express|) and mirroring on NASA's Planetary Data System Geosciences Node (\verb|http://pds-geosciences.wustl.edu/missions/mars_express/marsis.htm|). Simulations used in the identification of subsurface interfaces will be published in the same archives at a yet undefined date, but they are already available in the following public repository: \verb|https://doi.org/10.5281/zenodo.582651|.
\end{acknowledgments}

\end{article}

\end{document}